\documentclass[letterpaper]{article}
\usepackage{ijcai25}

\usepackage{times}
\usepackage{soul}
\usepackage{url}
\usepackage[hidelinks]{hyperref}
\usepackage[utf8]{inputenc}
\usepackage[small]{caption}
\usepackage{graphicx}
\usepackage{booktabs}

\usepackage[british]{babel}
\usepackage{relsize,setspace}
\usepackage{tikz,pgf}\usetikzlibrary{shapes,backgrounds}
\usepackage{amsmath,amssymb}
\usepackage[amsmath]{ntheorem}
\usepackage{cleveref}
\usepackage{multirow}
\usepackage{comment}
\usepackage{xspace}

\theoremseparator{.}
\theoremheaderfont{\normalfont\bfseries}
\theorembodyfont{\slshape}
\newtheorem{theorem}{Theorem}
\newtheorem{lemma}[theorem]{Lemma}
\newtheorem{proposition}[theorem]{Proposition}
\newtheorem{corollary}[theorem]{Corollary}

\theorembodyfont{\upshape}
\theoremsymbol{\ensuremath{\Diamond}}
\newtheorem{definition}{Definition}
\newtheorem{example}{Example}
\theoremstyle{nonumberplain}
\theoremheaderfont{\normalfont\bfseries}
\theorembodyfont{\slshape}

\theoremstyle{nonumberplain}
\theoremheaderfont{\normalfont\itshape}
\theoremsymbol{\ensuremath{\Box}}
\newtheorem{proof}{Proof}
\newtheorem{proofsketch}{Proof sketch} %
\theoremstyle{nonumberplain}
\theoremheaderfont{\normalfont\itshape}
\theoremsymbol{\ensuremath{\Diamond}}

\usepackage{ifthen}
\usepackage{environ}
\newif\iflong
\NewEnviron{shortproof}{\iflong\else\expandafter\begin{proof}\BODY\end{proof}\fi}
\NewEnviron{shortproofsketch}{\iflong\else\expandafter\begin{proofsketch}\BODY\end{proofsketch}\fi}
\NewEnviron{longproof}{\iflong\expandafter\begin{proof}\BODY\end{proof}\fi}

\usepackage{todonotes}

\newcommand{\N}{\mathbb{N}}

\newcommand{\R}{\mathbb{R}}

\newcommand{\set}[1]{\left\{#1\right\}}

\newcommand{\guard}{\ \middle\vert\ }

\newcommand{\tuple}[1]{\left(#1\right)}

\newcommand{\tand}{\;\text{and}\;}

\newcommand{\define}[1]{\emph{#1}}

\newcommand{\ordn}{\alpha}

\newcommand{\ordsucc}{\mathord{\ordn+1}}
\newcommand{\ordlim}{\beta}

\newcommand{\eqdef}{\mathbin{:=}}
\newcommand{\iffdef}{\mathbin{:\mkern-11mu\mathord{\iff}}}
\newcommand{\ebnfeq}{\mathrel{::=}}
\newcommand{\ebnfalt}{\mathbin{\mid}}

\newcommand{\citet}[1]{\citeauthor{#1}~[\citeyear{#1}]}

\newlength{\narrowgatherskip}
\newenvironment{narrowgather}[1][2pt]{%
    \setlength{\narrowgatherskip}{#1}%
    \vskip\narrowgatherskip\mbox{}\hfill\(}{\)\hfill\mbox{}\vskip\narrowgatherskip\noindent}

\newcommand{\flpcon}{f}
\newcommand{\Flpcon}{\mathcal{F}}
\newcommand{\Atoms}{\Pi}
\newcommand{\atom}{p}
\newcommand{\flpatom}{\Atoms}%
\newcommand{\fint}{I}
\newcommand{\fintalt}{J}
\newcommand{\uint}{U}
\newcommand{\lint}{L}
\newcommand{\Fint}{\mathcal{\fint}}
\newcommand{\fformeval}[2]{\hat{#1}\!\left(#2\right)}

\newcommand{\fnnformeval}[2]{\widehat{#1}\!\left(#2\right)}
\newcommand{\fnformeval}[3]{\fnnformeval{(#1,#2)}{#3}}
\newcommand{\flpcontf}[1]{#1^{\text{\raisebox{0.3ex}{\large$\mathord{\cdot}$}}}}
\newcommand{\flpnotf}{\mathord{\flpcontf{\flpnot}\mkern-1mu}}
\newcommand{\flpif}[1]{\overset{#1}{\gets}}
\newcommand{\flpnot}{\mathord{\sim}}

\newcommand{\flpand}{\land}
\newcommand{\flpor}{\lor}
\newcommand{\flpagg}{\mathbin{@}}
\newcommand{\flpweight}{\vartheta}
\newcommand{\flp}{\lp}

\newcommand{\flpr}{p \flpif{\flpweight}_i \body}
\newcommand{\flprRed}[1][\fint]{p \flpif{\flpweight}_i \body_{#1}}
\newcommand{\body}{\mathcal{B}}

\newcommand{\appInt}{\mathcal{C}}
\newcommand{\appint}{X}
\newcommand{\appintalt}{Y}
\newcommand{\rappint}[1][\appint]{#1^{\mathord{\triangleleft}}} %

\newcommand{\flpcontfStraccia}[1]{#1^{\text{\raisebox{0.3ex}{\large$\mathord{\cdot}\mathord{\cdot}$}}}}
\newcommand{\TpStraccia}[1][\flp]{\mathcal{S}_{#1}} %
\newcommand{\FpStraccia}[2][\flp]{F_{#1,#2}}
\newcommand{\SpStraccia}{s}
\newcommand{\WFStraccia}{\mathcal{AW}}
\newcommand{\pair}[2]{(#1,#2)}%

\newcommand{\flptext}{FLP\xspace}
\newcommand{\flpstext}{FLPs\xspace}

\newcommand{\mopo}{\mathit{mp}}

\newcommand{\lpif}{\gets}
\newcommand{\lpnot}{\mathord{\sim}}

\newcommand{\elp}{P}

\newcommand{\lset}{V}

\newcommand{\lleq}{\leqslant}

\newcommand{\llt}{<}

\newcommand{\lub}{\vee}
\newcommand{\glb}{\wedge}
\newcommand{\biglub}{\bigvee\!}
\newcommand{\bigglb}{\bigwedge\!}
\newcommand{\least}{0}
\newcommand{\grtst}{1}

\newcommand{\leastint}{\mathbf{\least}}
\newcommand{\grtstint}{\mathbf{\grtst}}

\newcommand{\fzi}[8]{%
    \ensuremath{%
        \scalebox{0.9}{%
            \(%
            \arraycolsep=1.5pt
            \renewcommand{\arraystretch}{0.8}
            \begin{array}{cccc}
                #2 & #4 & #6 & #8 \\
                #1 & #3 & #5 & #7
            \end{array}
            \)%
        }}%
}

\newcommand{\fst}[1]{#1_1}
\newcommand{\snd}[1]{#1_2}

\newcommand{\op}{O}
\newcommand{\ap}{\mathcal{A}}

\newcommand{\stable}[1]{#1^{\mathit{st}}}
\newcommand{\stap}{\stable{\ap}}

\newcommand{\complete}[1]{\mathord{c\stable{#1}}}
\newcommand{\lpoperator}[2]{#1_{#2}}
\newcommand{\lp}{P}
\newcommand{\lpop}[1][\lp]{\lpoperator{T}{#1}}

\newcommand{\lpap}[1][\lp]{\lpoperator{\mathcal{T}}{#1}}

\newcommand{\bil}[1][\lset]{#1^2}

\newcommand{\pleq}{\leq_p}
\newcommand{\plt}{<_p}
\newcommand{\plub}{\oplus}
\newcommand{\pglb}{\otimes}
\newcommand{\bigplub}{\bigoplus}
\newcommand{\tleq}{\leq_t}

\newcommand{\lfp}{\mathit{lfp}}

\newcommand{\ult}[1]{\mathcal{U}_#1}

\newcommand{\ultop}{\ult{\op}}
\newcommand{\ultlpop}{\ult{\flp}}

\tikzstyle{el}=[]
\tikzstyle{lt}=[draw,-,color=black]
\tikzstyle{ilt}=[draw,help lines,-,color=gray]
\tikzstyle{op}=[->,>=latex,draw,densely dotted,color=red,thick]
\tikzstyle{fp}=[loop above]
\tikzstyle{ap}=[->,>=latex,draw,loosely dashed,color=green,thick]
\tikzstyle{stap}=[->,>=latex,draw,densely dashed,color=violet,thick]
\tikzstyle{hdp}=[draw,circle,color=,minimum size=2pt,inner sep=1pt,outer sep=0pt]
\tikzstyle{hde}=[draw,-,color=black]
\tikzstyle{StracciaTP}=[->,>=latex,draw,loosely dashed,color=blue,thick]
\tikzstyle{StracciaSup}=[->,>=latex,draw,loosely dotted,color=teal,thick]
\tikzstyle{StracciaWP}=[->,>=latex,draw,densely dotted,color=orange,thick]

\newcommand\bilatticeEx{
  \node[el] (tpzero)                          {\fzi{0.0}{1.0}{0.0}{1.0}{0.0}{1.0}{0.0}{1.0}};
  \node[el] (tpone)    [above of=tpzero]      {\fzi{0.0}{1.0}{0.0}{1.0}{0.3}{0.6}{0.0}{1.0}};
  \node[el] (tpthree)  [above right of=tpone] {\fzi{0.3}{1.0}{0.0}{1.0}{0.3}{0.6}{0.0}{1.0}};
  \node[el] (awtpzero) [above left of=tpone]  {\fzi{0.0}{1.0}{0.0}{1.0}{0.3}{0.3}{0.0}{0.0}};
  \node[el] (awtpone) [above right of=awtpzero] {\fzi{0.3}{1.0}{0.0}{1.0}{0.3}{0.3}{0.0}{0.0}};
  \node[el] (awfour)  [above of=awtpone] {\fzi{0.3}{1.0}{0.0}{0.7}{0.3}{0.3}{0.0}{0.0}};
  \node[el] (spAWone) [below left of=awtpzero] {\fzi{0.0}{1.0}{0.0}{1.0}{0.0}{0.3}{0.0}{0.0}};
  \node[el] (spAWtwo) [above left of=spAWone] {\fzi{0.0}{1.0}{0.0}{0.7}{0.0}{0.3}{0.0}{0.0}};

  \node[el] (stable_candytwotpone) [above of=awfour, yshift=-3em]{\fzi{1.0}{1.0}{0.0}{0.0}{0.3}{0.3}{0.0}{0.0}};
  \node[el] (stable_candyawzero) [left of=stable_candytwotpone, xshift=-4em]{\fzi{0.3}{0.3}{0.0}{0.0}{0.3}{0.3}{0.0}{0.0}};
  \node[el] (stable_candyawone) [right of=stable_candytwotpone, xshift=4em]{\fzi{1.0}{1.0}{0.7}{0.7}{0.3}{0.3}{0.0}{0.0}};

  \path (tpzero) edge[ilt] (tpone);
  \path (tpone) edge[ilt] (tpthree);
  \path (awtpzero) edge[ilt] (tpone);
  \path (awtpzero) edge[ilt] (awtpone);
  \path (awtpone) edge[ilt] (tpthree);
  \path (awfour) edge[ilt] (awtpone);
  \path (spAWone) edge[ilt] (tpzero);
  \path (spAWone) edge[ilt] (awtpzero);
  \path (spAWtwo) edge[ilt] (awfour);
  \path (spAWtwo) edge[ilt] (spAWone);

  \path (tpzero) edge[StracciaTP, bend right=10] (tpone);
  \path (tpone) edge[StracciaTP, bend right=10] (tpthree);
  \path (tpthree) edge[StracciaTP, loop above] (tpthree);
  \path (tpzero) edge[StracciaSup, bend left=10] (spAWone);
  \path (awtpzero) edge[StracciaSup,bend right=10] (spAWone);
  \path (awtpone) edge[StracciaSup, bend right=10] (spAWtwo);
  \path (awfour) edge[StracciaSup, bend right=10] (spAWtwo);
  \path (tpzero) edge[stap, bend left] (awtpone);
  \path (awtpone) edge[stap, bend right] (awfour);
  \path (awfour) edge[stap, in=-10, out=10, loop] (awfour);
  \path (tpzero) edge[StracciaWP, bend left] (awtpzero);
  \path (awtpzero) edge[StracciaWP, bend left] (awtpone);
  \path (awtpone) edge[StracciaWP, bend left] (awfour);
  \path (awfour) edge[StracciaWP,in=190,out=170, loop ] (awfour);

  \path (stable_candytwotpone) edge[ilt] (stable_candyawone);
  \path (stable_candyawzero) edge[ilt] (stable_candytwotpone);
  \path (awfour) edge[ilt] (stable_candyawone);
  \path (awfour) edge[ilt] (stable_candytwotpone);
  \path (awfour) edge[ilt] (stable_candyawzero);

  \path (stable_candyawzero) edge[<->, >=latex,draw,densely dotted,color=orange,thick,bend left=10] (stable_candyawone);
  \path (stable_candyawone) edge[<->, >=latex,draw,dashed,color=blue,thick,bend right = 20 ] (stable_candyawzero);
  \path (stable_candytwotpone) edge[StracciaWP,in=-170,out=-150, loop] (stable_candytwotpone);
  \path (stable_candytwotpone) edge[StracciaTP,looseness=8, in=-10, out=-30, loop] (stable_candytwotpone);

  \node[el,rectangle, fill = white] (legend) [below of=stable_candyawone, xshift=1em, yshift=-14em] {
    \begin{tabular}{ll}
      \raisebox{0.5ex}{\tikz\path (0,0) edge[StracciaTP] (1,0);}  & $\lpap,\TpStraccia$ \\
      \raisebox{0.5ex}{\tikz\path (0,0) edge[StracciaSup] (1,0);} & $\SpStraccia_\flp$  \\
      \raisebox{0.5ex}{\tikz\path (0,0) edge[StracciaWP] (1,0);}  & $\WFStraccia_\elp$  \\
      \raisebox{0.5ex}{\tikz\path (0,0) edge[stap] (1,0);}        & $\stable{\lpap}$    \\
    \end{tabular}
  };

  \node[el,rectangle, fill = black!10] (program) [below of=legend, xshift=-3em, yshift=2em] {
    \begin{tabular}{l}
      $p \flpif{}_G \flpnot q \flpor_G r $          \\
      $q \flpif{}_G \flpnot p \flpor_G s $          \\
      $r \flpif{}_G 0.3 \flpor_G (s \flpand_G 0.6)$ \\
      $s \flpif{}_G s $                             \\
    \end{tabular}

  };

  \draw[->] (-7.5,-0.5) -- (5,-0.5) node[anchor=north west] {$\tleq$};
  \draw[->] (-7.5,-0.5) -- (-7.5,12) node[anchor=south east] {$\pleq$};

}

\longtrue

\title{Approximation Fixpoint Theory as a Unifying Framework for Fuzzy~Logic~Programming Semantics\iflong~(Extended Version)\fi}

\author{
    Pascal Kettmann$^1$
    \and
    Jesse Heyninck$^{2,3}$\and
    Hannes Strass$^{1,4}$\\
    \affiliations
    $^1$TU Dresden, Germany\\
    $^2$Open Universiteit, The Netherlands\\
    $^3$University of Cape Town, South Africa\\
    $^4$ScaDS.AI Center for Scalable Data Analytics and Artificial Intelligence, Dresden/Leipzig, Germany\\
    \emails
    \{pascal.kettmann, hannes.strass\}@tu-dresden.de,
    jesse.heyninck@ou.nl
}

\hypersetup{
    pdftitle={Approximation Fixpoint Theory as a Unifying Framework for Fuzzy Logic Programming Semantics},
    pdfauthor={Pascal Kettmann, Jesse Heyninck, Hannes Strass}
}

\begin{document}

\maketitle

\begin{abstract}
  Fuzzy logic programming is an established approach for reasoning under uncertainty.
Several semantics from classical, two-valued logic programming have been generalized to the case of fuzzy logic programs.
In this paper, we show that two of the most prominent classical semantics, namely the stable model and the well-founded semantics, can be reconstructed within the general framework of approximation fixpoint theory (AFT).

This not only widens the scope of AFT from two- to many-valued logics, but allows a wide range of existing AFT results to be applied to fuzzy logic programming.
As first examples of such applications, we clarify the formal relationship between existing semantics, generalize the notion of stratification from classical to fuzzy logic programs, and devise “more precise” variants of the semantics.

\end{abstract}

\section{Introduction}

Logic programs~\cite{KowalskiK71,Lloyd87} are an established language not only for programming, but also for knowledge representation and reasoning.
However, in classical logic programs, the dichotomy of allowing propositions to only be either true or false can
limit expressivity.
More precisely, logic programs under two-valued semantics cannot easily represent knowledge that is imprecise, vague, or uncertain.
To address this limitation, \citet{lee72} proposed to generalize logic programs to enable the use of \emph{fuzzy logic} in its semantics, which among other things means allowing to use truth values from the real unit interval $[0,1]$.

This initial proposal of \emph{fuzzy logic programs} (\flpstext) has led to a considerable body of research
that continues growing to this day \cite{Shapiro83,vEmden86,Subrahmanian87,Vojtas01,MedinaOV01,LoyerS02,loyer2003approximate,LoyerS09,cornejo2018syntax,Cornejo2020extended}, with an increasing interest in frameworks that abstract away from concrete sets of truth values and instead require only some algebraic properties on them, thus also encompassing the classical, two-valued version as a special case.
However, there is no uniform approach to syntax, e.g.\ what kinds of connectives are allowed in rule bodies, and also no universally accepted “standard” semantics.
Instead, several proposals exist, e.g.\ the approximate well-founded semantics by \citet{loyer2009approximate} or the stable model semantics by \citet{cornejo2018syntax}, %
but it is not clear how they are related.

In the present paper, we address some of these open questions within the general algebraic framework of \emph{approximation fixpoint theory} (AFT).
For classical logic programming~\cite{vanEmdenK76}, it is known that the (standard) least-model semantics of definite logic programs can be defined using an operator on interpretations.
In a series of papers, \citeauthor{denecker00approximations} demonstrated that similar constructions are possible for a range of knowledge representation formalisms, obtaining several profound results that uncovered fundamental similarities between semantics of different knowledge representation languages~\cite{denecker00approximations,denecker03uniformsemantic,denecker04ultimate}.
Today, approximation fixpoint theory stands as a unifying framework for studying fixpoint-based semantics, in particular in the context of languages allowing for non-monotonicity.

More concretely, in this paper we show that several fuzzy logic programming semantics from the literature can be incorporated into the framework of approximation fixpoint theory:
\begin{enumerate}
    \item the approximate well-founded semantics by \citet{loyer2009approximate}, a generalization of the well-founded semantics for classical logic programming \cite{GelderRS91}; and
    \item the stable model semantics by \citet{cornejo2018syntax}, a generalization of the classical stable model semantics \cite{GelfondL88}.
\end{enumerate}
In fact, we will show that both semantics can be obtained \emph{by a single operator} using standard constructions from approximation fixpoint theory.
As direct results, we obtain how the two semantics are related, and that we can also apply other known techniques from AFT directly to fuzzy logic programming, which we demonstrate by generalizing stratification and ultimate approximation.
We are not aware of any existing reconstructions of fuzzy logic programming semantics within approximation fixpoint theory;
perhaps the closest related work is that on weighted abstract dialectical frameworks~\cite{BrewkaSWW18} (that are conceptually similar to fuzzy logic programs, but with a different area of application) with an existing treatment in AFT~\cite{Bogaerts19}.

The paper proceeds as follows:
We next give necessary background on approximation fixpoint theory (insofar it is relevant for logic programming) and fuzzy logic programming.
Afterwards, we define the concrete syntax of fuzzy logic programs and define the operator that we subsequently use to reconstruct the well-founded semantics (\Cref{sec:wf}) and the stable model semantics (\Cref{sec:sm}).
In \Cref{sec:aft4flp}, we showcase some applications of AFT to fuzzy logic programming, reaping the immediate benefits of our reconstruction.
Finally, we discuss some avenues for further work and conclude.

\section{Background}

In this section, we briefly introduce AFT, cover basic fuzzy logic principles, and give an overview of (truth-functional) fuzzy logic programming, focusing on the operator defined by \citet{Vojtas01} and many-valued modus ponens.

\subsection{Approximation Fixpoint Theory}\label{subsec:aft}

A particular way of defining semantics for programming languages is to associate a program $P$ with a \emph{transformation} operator $T_P$ that transforms a given input-output relation $R$ into an updated relation $T_P(R)$.
In this view, the semantics of program $P$ is given by the least fixpoint of the operator $T_P$.
Van Emden and Kowalski~[\citeyear{vanEmdenK76}] applied this way of defining semantics to the field of logic programming, developing the first one-step logical consequence operator and employing its monotonicity in conjunction with \citeauthor{tarski55fixpoint}'s~fixpoint theorem~[\citeyear{tarski55fixpoint}] to define the intended semantics of definite logic programs with potentially recursive predicate definitions.

More formally, an \define{operator} is a (total) function $\op\colon\lset\to\lset$ on a set $\lset$.
We are typically interested in operators on sets $\lset$ with an internal structure:
At the minimum, we require $(\lset,\lleq)$ to be a partially ordered set.
More often, $(\lset,\lleq)$ will be a \define{complete lattice}, that is, such that every subset $S\subseteq\lset$ has a
\define{least upper bound (lub)} $\biglub S\in\lset$ and
\define{greatest lower bound (glb)} $\bigglb S\in\lset$;
this entails that $(\lset,\lleq)$ has both a least element $\least=\bigglb\lset=\biglub\emptyset$ and a greatest element $\grtst=\biglub\lset=\bigglb\emptyset$.
An operator $\op\colon\lset\to\lset$ is
\define{monotone} iff $x\lleq y$ implies $\op(x)\lleq\op(y)$, and
\define{antimonotone} iff $x\lleq y$ implies $\op(y)\lleq\op(x)$.
An element $x\in\lset$ with $\op(x)=x$ is a \define{fixpoint} of $\op$;
if $\op(x)\lleq x$ then $x$ is a \define{pre-fixpoint} of $\op$;
if $x\lleq\op(x)$ then $x$ is a \define{post-fixpoint} of $\op$.
A fundamental result in lattice theory, \citeauthor{tarski55fixpoint}'s~fixpoint theorem~[\citeyear{tarski55fixpoint}], states that whenever $\op$ is a monotone operator on a complete lattice $(\lset,\lleq)$, the set $\set{x\in\lset \guard \op(x)=x }$ of its fixpoints forms a complete lattice itself, and so has a least element $\lfp(\op)$.
While this result is immensely useful, it is not constructive;
however, constructive versions exist and tell us how to actually construct least fixpoints~\cite{markowsky1976chain,cousot1979constructive}:
To this end, one defines, for any $x\in\lset$, the iterated (transfinite) application of $\op$ to $x$ via
$\op^{0}(x)\eqdef x$,
$\op^{\ordsucc}(x)\eqdef\op(\op^{\ordn}(x))$ for successor ordinals, and
$\op^{\ordlim}(x)\eqdef\biglub\set{\op^{\ordn}(x) \guard \ordn<\ordlim}$ for limit ordinals.
It is then known that (for monotone $\op$) there exists an ordinal $\ordn$ such that $\op^{\ordn}(\least)=\lfp(\op)$~\cite{bourbaki49theoreme}.

In order to apply \citeauthor{tarski55fixpoint}'s~fixpoint theorem~[\citeyear{tarski55fixpoint}] (or its constructive versions), it is thus necessary to have an operator that is monotone, a condition that is no longer fulfilled when extending the syntax of logic programs from \emph{definite} to \emph{normal} logic programs, i.e., when allowing “negation as failure” to occur in bodies of rules.
To still have a useful fixpoint theory in the case of non-monotone operators, \citet{denecker00approximations} fundamentally generalized the theory underlying the fixpoint-based approach to semantics, thereby founding what is now known as \emph{approximation fixpoint theory}.
Its underlying idea is that when an operator of interest does not have properties that guarantee the existence of fixpoints, then this operator can be approximated in a more fine-grained algebraic structure where fixpoint existence can be guaranteed.

Formally -- following ideas of \citet{belnap1977useful}, \citet{Ginsberg88a}, and \citet{fitting02fixpointsemantics} --, \citet{denecker00approximations} moved from a complete lattice $(\lset,\lleq)$ to its associated \define{bilattice} on the set $\bil\eqdef\lset\times\lset$.
So where elements of $\lset$ correspond to candidates for the semantics (interpretations), the {pairs} contained in $\bil$ correspond to \emph{approximations} of such candidates.
More technically, a pair $(x,y)\in\bil$ approximates all $z\in\lset$ with $x\lleq z\lleq y$.
A pair $(x,y)\in\bil$ is called \define{consistent} iff $x\lleq y$, in other words, if the interval $[x,y]\eqdef\set{z\in\lset\guard x\lleq z\lleq y}$ is non-empty;
a pair of the form $(x,x)$ is \define{exact}.
There are two natural orderings on $\bil$:
\begin{itemize}
    \item $(x_1,y_1)\tleq(x_2,y_2)\iffdef x_1\lleq x_2 \tand y_1\lleq y_2$, the \define{truth ordering} extending the (truth) lattice ordering $\lleq$;
    \item $(x_1,y_1)\pleq(x_2,y_2)\iffdef x_1\lleq x_2 \tand y_2\lleq y_1$, the \define{precision ordering} comparing intervals by precision of approximation.
          The greatest lower bound induced by $\pleq$ is denoted by $\pglb$ and
          the least upper bound by $\plub$, where we have
          $(x_1,y_1)\pglb(x_2,y_2)=(x_1\glb x_2,y_1\lub y_2)$ and
          $(x_1,y_1)\plub(x_2,y_2)=(x_1\lub x_2,y_1\glb y_2)$.
\end{itemize}
Just as pairs of $\bil$ approximate elements of $\lset$, the main idea of \citet{denecker00approximations} was that operators $\op\colon\lset\to\lset$ can be approximated by operators on $\bil$:
An \define{approximator} is an operator $\ap\colon\bil\to\bil$ that is $\pleq$-monotone and that maps exact pairs to exact pairs.
By virtue of the latter condition, we say that $\ap$ \define{approximates} the operator $\op\colon\lset\to\lset$ with $\op(x)=\fst{\ap(x,x)}$ (where $\fst{(x,y)}=x$ and $\snd{(x,y)}=y$);
by virtue of the first condition, $\ap$ is guaranteed to possess ($\pleq$-least) fixpoints.
An approximator $\ap$ is \define{symmetric} iff for all $(x,y)\in\bil$, we have $\snd{\ap(x,y)}=\fst{\ap(y,x)}$;
thus symmetry allows to specify $\ap$ by giving only $\fst{\ap(\cdot,\cdot)}$.

The main contribution of \citet{denecker00approximations} was the association of the \define{stable approximator $\stap$} to an approximator $\ap$:
For a complete lattice \mbox{$(\lset,\lleq)$} and an approximator \mbox{$\ap\colon\bil\to\bil$}, define the
\define{stable approximator for $\ap$} as \mbox{$\stap\colon\bil\to\bil$} by
\mbox{$\stap(x,y) \eqdef (\lfp(\fst{\ap(\cdot,y)}),\lfp(\snd{\ap(x,\cdot)}))$}, where $\fst{\ap(\cdot,y)}$ (resp.\ $\snd{\ap(x,\cdot)}$) denotes the operator that maps any $z\in\lset$ to $\fst{\ap(z,y)}$ (resp.\ to $\snd{\ap(x,z)}$).
Among other results, \citet{denecker00approximations} also showed that this construction is well-defined because
both relevant operators
\mbox{$\fst{\ap(\cdot,y)}\colon\lset\to\lset$} and
\mbox{$\snd{\ap(x,\cdot)}\colon\lset\to\lset$}
are $\lleq$-monotone (as $\ap$ is an approximator) and thus possess least fixpoints.
Furthermore, any approximator maps consistent pairs to consistent pairs, and the stable approximator $\stap$ maps consistent post-fixpoints of $\ap$ to consistent pairs.
Consequently, not only is $\lfp(\ap)$ consistent (and called the \define{Kripke-Kleene fixpoint} of $\ap$), but so is $\lfp(\stap)$, the \define{well-founded fixpoint} of $\ap$.
Moreover, a fixpoint $(x,y)$ of $\stap$, called a \define{stable fixpoint} of $\ap$, is always a $\lleq$-minimal fixpoint of $\op$ (when $\ap$ approximates $\op$).
The term “stable” is not coincidental: When using the definite logic program transformation operator $T_P$ by \citet{vanEmdenK76} as $\op$ and defining a suitable approximator~$\ap$ (see \Cref{sec:approximator} for a generalization of such an approximator), then the exact fixpoints of $\stap$ correspond one-to-one to the stable models of the logic program~$P$ \cite{denecker00approximations}.

\subsection{Fuzzy Logic Programming}

Let us first define the syntax of \emph{positive} fuzzy formulas, a generalization of the conjunctions of literals that appear in classical definite logic programs.
Given a complete lattice $(\lset,\lleq)$ of truth values,
a set $\Atoms$ of propositional atoms, and
a family $\Flpcon$ of fuzzy connectives (where each $f\in\Flpcon$ has an associated \define{arity} $n\in\N$ which we denote by $f^{(n)}$ when important),
we designate a finitely representable subset $\lset'\subseteq V$ of truth values that are allowed to be explicitly mentioned in programs,%
\footnote{Allowing truth values to occur explicitly in formulas is with the intended usage of being able to represent \emph{some} truth values in rule bodies as we do in \Cref{exm:main}.
This does not necessarily work for arbitrary sets of truth values (e.g.\ it does not work for $\lset=[0,1]\subseteq\R$ by sheer cardinality) -- in those cases we want to allow to restrict the syntax to a meaningful subset of truth values that can be finitely represented, e.g.\ the rational unit interval $\lset'=[0,1]\cap\mathbb{Q}$.}
and define the syntax of a \define{positive fuzzy formula} via
\begin{equation}
  \varphi \ebnfeq c \ebnfalt p \ebnfalt \varphi \flpand \varphi \ebnfalt \varphi \flpor \varphi \ebnfalt \flpagg(\varphi, \dots ,\varphi)
  \label{eq:positive-formula}
\end{equation}
where $c\in\lset'$, $p\in\Atoms$, and $\flpand,\flpor,\flpagg \in \Flpcon$, to which we refer as conjunctors, disjunctors, and aggregators, respectively.
A \define{positive logic program rule} is an expression of the form
\begin{equation}
  p \flpif{\flpweight} \body
  \label{eq:flprule}
\end{equation}
where $p\in\Atoms$, $\flpweight\in\lset'$, $\flpif{}$ is an implicator and $\mathcal{B}$ is a positive fuzzy formula. %
The atom $p$ is called the \define{head} of the rule, $\mathcal{B}$ is called the \define{body} of the rule and $\flpweight$ its \define{weight} (whenever $\flpweight=\grtst$, we will write a rule as $p \flpif{} \body$).
A \define{positive fuzzy logic program} $\flp$ is then a finite set of rules~(\ref{eq:flprule}).

Every connective \mbox{$\flpcon^{(n)}\in\Flpcon$} has an associated truth function \mbox{$\flpcontf{\flpcon}\colon\lset^n\to\lset$} on the complete lattice $(\lset, \lleq)$ of truth values.
The truth functions of the (positive) connectives in our language are assumed to satisfy the following properties:
\begin{itemize}
  \item for each $\flpcon \in \Flpcon$, $\flpcontf{\flpcon}$ is $\lleq$-monotone in each argument;
  \item for each conjunctor \mbox{$\mathord{\flpand}\in\Flpcon$}, its truth function~$\flpcontf{\flpand}$ \mbox{satisfies\!}
        \mbox{$\atom_1\flpcontf{\flpand}\atom_2\lleq \atom_1$},
        \mbox{$\atom_1\flpcontf{\flpand}\atom_2\lleq \atom_2$}, and
        \mbox{$\atom\flpcontf{\flpand}\grtst=\atom=\grtst\flpcontf{\flpand}\atom$};
  \item for each disjunctor \mbox{$\mathord{\flpor}\in\Flpcon$}, its truth function $\flpcontf{\flpor}$ satisfies
        \mbox{$\atom_1\lleq\atom_1\flpcontf{\flpor}\atom_2$},
        \mbox{$\atom_2\lleq\atom_1\flpcontf{\flpor}\atom_2$}, and
        \mbox{$\atom\flpcontf{\flpor}\least=\atom=\least\flpcontf{\flpor}\atom$};
  \item for each implicator $\mathord{\flpif{}}\in\Flpcon$, $\flpcontf{\flpif{}}$ is monotone in the first argument and antimonotone in the second argument.
\end{itemize}
If $\flpcontf{\flpand}$ is also associative and commutative and \mbox{$\lset = [0,1]\subseteq\R$}, then $\flpcontf{\flpand}$ is called a \define{t-norm}.
However, we do not make any assumptions on associativity, commutativity, or continuity of truth functions in this paper.
An example of truth functions are Gödel's, with
$x \flpcontf{\flpand_G}y = \min\{x, y\}$,
$x \flpcontf{\flpor_G}y = \max\{x, y\}$, and
$x \flpcontf{\flpif{}_G}y = x$ if $y > x$ and $x\flpcontf{\flpif{}_G} y= 1$ otherwise.

For a complete lattice $(\lset, \lleq)$ of truth values and $\Atoms$ a set of atoms, %
an \emph{interpretation} is a total function $\fint \colon \Atoms \to \lset$. %
The set of all interpretations is denoted by $\Fint$.
It can be shown (and is known) that $(\Fint,\leq)$ again is a complete lattice, where $\fint\leq\fintalt$ iff $\fint(\atom)\lleq\fintalt(\atom)$ for all $\atom\in\Atoms$.
We denote the least and greatest elements of this lattice by $\leastint\eqdef\set{\atom\mapsto\least\guard\atom\in\Atoms}$ and $\grtstint\eqdef\set{\atom\mapsto\grtst\guard\atom\in\Atoms}$, respectively.
Later, we will work on the associated bilattice $\bil[\Fint]$ with $\pleq$-least element $(\leastint,\grtstint)$;
glb $\pglb$ and lub $\plub$ carry over from $\bil$ to $\bil[\Fint]$ in a pointwise manner, e.g.~$((\lint,\uint)\pglb(\lint',\uint'))(\atom)\eqdef (\lint,\uint)(\atom)\pglb(\lint',\uint')(\atom)$.

We work with truth-functional logic in a narrow sense, whence the truth value of a formula is uniquely determined by the truth value of its constituents.
Given an interpretation $\fint$, we thus extend the interpretation to arbitrary formulas $\varphi$ via structural induction, and denote the truth value of $\varphi$ under $\fint$ by $\fformeval{\fint}{\varphi}$.
More precisely, for connective \mbox{$f^{(n)}\in\Flpcon$}, we define \mbox{$\fformeval{\fint}{f(\varphi_1,\ldots,\varphi_n)}\eqdef \flpcontf{f}(\fformeval{\fint}{\varphi_1},\ldots,\fformeval{\fint}{\varphi_n})$}.
For example, for Gödel logic we obtain $\fformeval{\fint}{x \flpand_G y} = \fint(x)\flpcontf{\flpand_G}\fint(y)$.

We next define how the truth value $\fformeval{\fint}{\body}$ of a body of a rule is propagated towards the head $\atom$.
The inference rule proposed by \citet{Vojtas01} for fuzzy logic programming is based on many-valued modus ponens (\citet{Hajek1998} gives details):
\begin{equation*}
  \tag{\text{MP}}
  \frac{p \flpif{} \varphi\quad \varphi}{p}
\end{equation*}
(MP) should infer a truth value $z$ of the conclusion $p$ when given
a truth value $x$ of $\varphi$ and
a truth value $y=z\flpcontf{\flpif{}}x$ of the implication $p\flpif{}\varphi$.
Let $\mopo$ denote a candidate function, i.e.\ a way to compute $\mopo(x,y)=z$.
To ensure soundness of the inference, $\mopo$ should be monotone in both of its arguments (increasing truth values of the premises lead to higher truth in the conclusion) and satisfy the boundary conditions $\mopo(0,1)=\mopo(1,0)=0$ and $\mopo(1,1)=1$.
These are the properties of a truth function of a conjunctor $\flpand$.
If in addition the truth value of $z \flpcontf{\flpif{}} x$ is as large as possible, we get
\begin{equation}
  x \flpcontf{\flpand} y \lleq z \text{ iff } y \lleq z \flpcontf{\flpif{}} x.
  \label{eq:adjoint}
\end{equation}
A pair $(\flpcontf{\flpif{}}, \flpcontf{\flpand})$ that satisfies \eqref{eq:adjoint} for all $x,y,z \in \lset$ of a partially ordered set $(\lset,\lleq)$ is then called an \define{adjoint pair}~\cite{MedinaOV01}.
In order to designate adjoint pairs, we will write them with the same index, i.e.\ as $(\flpif{}_i,\flpand_i)$.
Intuitively, adjoint pairs guarantee correctness of inference whenever we pair “matching” implicators and conjunctors, e.g.~$(\flpif{}_G, \flpand_G)$. %

Given an interpretation \mbox{$\fint \in \Fint$}, a weighted rule \mbox{$\flpr$} is \emph{satisfied} by $\fint$ iff \mbox{$\flpweight \lleq \fint(p)\flpcontf{\flpif{}_i} \fformeval{\fint}{\mathcal{B}}$}.
An interpretation $\fint$ is a \emph{model} of $\flp$ iff it satisfies all rules in $\flp$.
Intuitively, $\fint$ satisfies a rule if it makes the rule “at least as true” as its weight $\flpweight$.
By \eqref{eq:adjoint}, this translates to:
the head is “at least as true” as the body limited by the rule's weight, or formally $\flpweight \flpcontf{\flpand_i} \fformeval{\fint}{\mathcal{B}} \lleq \fint(p)$.

\citet{Vojtas01} and \citet{MedinaOV01} define a fuzzy generalization of the transformation operator given by \citet{vanEmdenK76}:
For a fuzzy logic program $\flp$, the immediate consequence operator $\lpop \colon \Fint \to \Fint$ is %
defined by
\begin{equation}
  \lpop(\fint)(\atom) = \biglub \set{\flpweight\flpcontf{\flpand_i}\fformeval{\fint}{\body} \guard \flpr \in \flp}
  \label{eq:}
\end{equation}
The semantics of a positive \flptext $\flp$ can then be characterized by the pre-fixpoints of $\lpop$~\cite[Theorem~2.2]{Vojtas01}, i.e.\ an interpretation \mbox{$\fint \in \Fint$} is a model of $\flp$ iff \mbox{$\lpop(\fint)\leq \fint$}.
This generalizes a classical result by \citet{vanEmdenK76}.

\begin{example}
  \label{exm:simple}
  Consider the positive fuzzy logic program \mbox{$\flp = \set{r \flpif{}_G 0.3 \flpor_G (s \flpand_G 0.6), s \flpif{}_G s}$} over %
  $([0,1],\lleq)$.
  Any interpretation $\fint$ with $0.3 \lleq \fint(r)$ %
  is a model of $\flp$.
  The \emph{least} model of $\flp$ can be obtained as the least fixpoint of $\lpop$ via
  transfinitely iterating $\lpop$ on $\leastint$, leading to the interpretation
  $\lpop(\leastint) = \set{r\mapsto 0.3, s\mapsto 0} = \fint_1 = \lpop(\fint_1)=\lfp(\lpop)$.%
\end{example}

\section{An Approximator for FLPs}
\label{sec:approximator}

The main challenge in defining semantics for (fuzzy) logic programs lies in the treatment of “negation as failure”.
Syntactically, we now move to \define{normal} fuzzy logic programs, that is, we extend the syntax of positive fuzzy formulas by \emph{default negation} $\flpnot$, a unary connective of which we only require that its truth function \mbox{$\flpnotf\colon\lset\to\lset$} is an antimonotone involution~\cite{Ovchinnikov1983fuzzynegation}.
($\flpnotf$ being an involution means that \mbox{$\flpnotf\flpnotf v=v$} for all \mbox{$v\in\lset$};
it follows that \mbox{$\flpnotf\least=\grtst$} and \mbox{$\flpnotf\grtst=\least$} whence $\flpnot$ is a generalization of two-valued negation.)
For \mbox{$\lset=[0,1] \subseteq\R$}, the typical choice is \mbox{$\flpnotf x=1-x$}, which we will use in all subsequent examples.

Formally, a \define{normal fuzzy formula} is of the form
\begin{equation}
  \varphi \ebnfeq c \ebnfalt \atom \ebnfalt \flpnot\atom \ebnfalt \varphi \flpand \varphi \ebnfalt \varphi \flpor \varphi \ebnfalt \flpagg(\varphi, \dots ,\varphi) %
  \label{eq:normal-formula}
\end{equation}
where $c\in\lset'$, $p\in\Atoms$, and $\flpand,\flpor,\flpagg \in \Flpcon$ as before.
A \define{normal fuzzy logic program} is then a finite set of normal fuzzy rules, and the remaining notions involving “normal” carry over from “positive” as expected.
In particular, to define the evaluation of a normal fuzzy formula $\varphi$ under interpretation \mbox{$\fint\in\Fint$}, we have the additional inductive case \mbox{$\fformeval{I}{\flpnot\varphi}=\flpnotf\fformeval{\fint}{\varphi}$}.

The one-step consequence operator $\lpop$ \cite{Vojtas01,MedinaOV01} is likewise extended to the case of normal \flpstext, with models of $\flp$ and pre-fixpoints of $\lpop$ still being in one-to-one correspondence.
However, the resulting operator $\lpop$ is no longer $\leq$-monotone and so cannot directly be used to define a least-model semantics for normal programs.
\begin{example}
  \label{exm:main}
  Consider the following normal fuzzy logic program $\flp$ over $\lset=[0,1]$ (due to \citet{loyer2009approximate}):

  \begin{tabular}{p{4cm} p{4cm}}
    $ p \flpif{}_G \flpnot q \flpor_G r, $           & $q \flpif{}_G \flpnot p \flpor_G s, $ \\
    $ r \flpif{}_G 0.3 \flpor_G (s \flpand_G 0.6),$ & $s\flpif{}_G s.$
  \end{tabular}
  Any fuzzy interpretation $\fint$ with
  $0.3 \lleq \fint(r) \lleq \fint(p)$,
  $\fint(s) \lleq \fint(q)$, and
  $1 - \fint(p) \lleq \fint(q)$ is a model of $\flp$.
  Whenever $\fint(p)+\fint(q)=1$,
  $0.3 \lleq \fint(p)$,
  $\fint(r)=0.3$, and $\fint(s)=0$, then $\fint$ is also a \emph{minimal} model.
  Consequently, $\flp$ has an infinite number of minimal models.
\end{example}
Approximation fixpoint theory fortunately provides an elegant solution to this problem --
all we have to do is define an approximator to $\lpop$, and AFT does the rest.

\begin{definition}[Approximator]
  \label{def:approximator}
  For a normal \flptext $\flp$ over complete lattice $(\lset,\lleq)$, we associate the symmetric approximator $\lpap\colon\bil[\Fint]\to\bil[\Fint]$ with
  \[
    \fst{\lpap(\lint,\uint)}(\atom)=
    \biglub \set{\flpweight\flpcontf{\flpand_i}\fnformeval{\lint}{\uint}{\body} \guard \flpr \in \flp}
  \]
  where $\fnformeval{\lint}{\uint}{\body}$, the evaluation of $\body$ under interpretation pair $(\lint,\uint)\in\bil[\Fint]$, is defined via structural induction with notable base cases
  $\fnformeval{\lint}{\uint}{\atom}\eqdef\lint(\atom)$ and
  $\fnformeval{\lint}{\uint}{\flpnot\atom}\eqdef\flpnotf\uint(\atom)$,
  and straightforward inductive case $\fnformeval{\lint}{\uint}{f(\varphi_1,\ldots,\varphi_n)}\eqdef\flpcontf{f}\!\left(\fnformeval{\lint}{\uint}{\varphi_1},\ldots,\fnformeval{\lint}{\uint}{\varphi_n}\right)$.
\end{definition}
Thus $\lpap(\lint,\uint)=\tuple{\fst{\lpap(\lint,\uint)},\fst{\lpap(\uint,\lint)}}$ gives the overall approximator.
\Cref{fig:bilattice2}~(p.~\pageref{fig:bilattice2}) shows how the approximator maps (pairs of) interpretations to others for \Cref{exm:main}.

It is easy to see that $\lpap$ is well-defined to serve its purpose:
\begin{proposition}
  \label{prop:approximator}
  $\lpap$ approximates $\lpop$ and is $\pleq$-monotone.
  \begin{longproof}
    \begin{description}
      \item[\normalfont$\lpap$ approximates $\lpop$:]
            For any $\fint\in\Fint$, by symmetry of $\lpap$ we have
            $\fst{\lpap(\fint,\fint)}=\snd{\lpap(\fint,\fint)}$ with
            $$\fst{\lpap(\fint,\fint)}(\atom)=
              \biglub \set{\flpweight\flpcontf{\flpand_i}\fnformeval{\fint}{\fint}{\body} \guard \flpr \in \flp}$$
            whence the claim follows because $\fnformeval{\fint}{\fint}{\body}=\fformeval{\fint}{\body}$.
      \item[\normalfont$\lpap$ is $\pleq$-monotone:]
            Consider $(\lint_1,\uint_1)\pleq(\lint_2,\uint_2)$.
            For any $\atom\in\Atoms$, we have $\uint_2(\atom)\lleq\uint_1(\atom)$, whence
            $\flpnotf\uint_1(\atom)\lleq\flpnotf\uint_2(\atom)$ (because $\flpnotf$ is antimonotone).
            Since additionally $\lint_1(\atom)\lleq\lint_2(\atom)$, we obtain
            $\fnformeval{\lint_1}{\uint_1}{\atom}\lleq\fnformeval{\lint_2}{\uint_2}{\atom}$.
            By induction, (for every $f\in\Flpcon$, $\flpcontf{f}$ is pointwise $\lleq$-monotone) we can then show
            $\fnformeval{\lint_1}{\uint_1}{\varphi}\lleq\fnformeval{\lint_2}{\uint_2}{\varphi}$
            for every normal fuzzy formula $\varphi$.
            Now consider $\atom\in\Atoms$ and the set $\body_\atom=\set{\flpweight\flpand_i\body \guard \flpr\in\flp }$.
            Denoting $\fnformeval{\lint}{\uint}{\body_\atom}\eqdef\set{\fnformeval{\lint}{\uint}{\varphi}\guard\varphi\in\body_\atom}$, it remains to show
            $\biglub\fnformeval{\lint_1}{\uint_1}{\body_\atom}\lleq\biglub\fnformeval{\lint_2}{\uint_2}{\body_\atom}$ which shows
            $\fst{\lpap(\lint_1,\uint_1)}\lleq\fst{\lpap(\lint_2,\uint_2)}$ and by symmetry of $\lpap$ proves the claim.
            For any $\varphi\in\body_\atom$, we have $\fnformeval{\lint_1}{\uint_1}{\varphi}\lleq\fnformeval{\lint_2}{\uint_2}{\varphi}\lleq\biglub\fnformeval{\lint_2}{\uint_2}{\body_\atom}$, thus $\biglub\fnformeval{\lint_2}{\uint_2}{\body_\atom}$ is an upper bound of $\fnformeval{\lint_1}{\uint_1}{\body_\atom}$.
            Since $\biglub\fnformeval{\lint_1}{\uint_1}{\body_\atom}$ is the least upper bound of $\fnformeval{\lint_1}{\uint_1}{\body_\atom}$, the claim follows.
            \hfill$\square$ %
    \end{description}
  \end{longproof}
\end{proposition}

Thus $\lpap$ also approximates the operator by \citet{Vojtas01} and generalizes its semantics.
Additionally, using the standard definitions of approximation fixpoint theory, \Cref{def:approximator} also indirectly yields the \define{well-founded semantics} of $\flp$ via $\lfp(\stable{\lpap})$, as well as the \define{stable model semantics} via the exact fixpoints of $\stable{\lpap}$.
In the remainder of the paper, we will relate these semantics (as obtained by AFT directly) to similar semantics that have been manually defined in the literature.

\section{Well-Founded Semantics}
\label{sec:wf}

In this section we show that AFT captures the well-founded semantics for fuzzy logic programs as defined by \citet{LoyerS09}.
The first (minor) technical hurdle is that where AFT uses pairs $(\lint,\uint)$ of fuzzy interpretations, \citet{LoyerS09} use interpretations that assign a \emph{pair} of truth values to each atom \mbox{$\atom\in\Atoms$}.
More formally, an \define{approximate interpretation} is a total function \mbox{$\appint\colon\Atoms\to\lset\times\lset$};
the set of all approximate interpretations is denoted by $\appInt$.
\citet{LoyerS09} then extend approximate interpretations to normal fuzzy formulas:
Firstly, they extend the truth functions of connectives $f^{(n)}\in\Flpcon$ from $\lset$ to $\lset\times\lset$ via
$\flpcontfStraccia{f}(\pair{\ell_1}{u_1},\ldots,\pair{\ell_n}{u_n})\eqdef \pair{\flpcontf{f}(\ell_1,\ldots,\ell_n)}{\flpcontf{f}(u_1,\ldots,u_n)}$, and
set $\flpcontfStraccia{\flpnot}(\pair{\ell}{u})\eqdef \pair{\flpnotf u}{\flpnotf \ell}$;
secondly, they define $\fformeval{\appint}{\varphi}$ by structural induction on formulas $\varphi$ (including default negation $\flpnot$), with base cases
$\fformeval{\appint}{\atom}=\appint(\atom)$ and $\fformeval{\appint}{c}=\pair{c}{c}$ for $c\in\lset'$.
\citet{LoyerS09} then define an operator on approximate interpretations via
\(
\TpStraccia(\appint)(\atom)=\fformeval{\appint}{\body_\atom}
\)
where they assume that every $\atom\in\Atoms$ has a unique rule $\atom\lpif\body_\atom\in\flp$ (which is not a restriction because several rules for $\atom$ can be joined with disjunction $\flpor$%
).
It is clear that the set $\appInt$ of approximate interpretations and the set $\bil[\Fint]$ of pairs $(\lint,\uint)$ of fuzzy interpretations are isomorphic (whence $\pglb$ and $\plub$ carry over to $\appInt$);
for a given $\appint\in\appInt$, we use $\zeta(\appint)$ to denote $(\lint,\uint)$ such that $\appint(p)=\pair{\lint(p)}{\uint(p)}$ for all $p\in\Atoms$.
As our first result in this section, it then follows that modulo this isomorphism on the underlying structures, the operator $\TpStraccia$ by \citeauthor{LoyerS09} [\citeyear{LoyerS09}] and our approximator $\lpap$ of \Cref{def:approximator} coincide.

\begin{lemma}
    \label{lem:straccia:operators-isomorphic}
    For any \flptext $\flp$ and any approximate interpretation $\appint$ over $\Atoms$, we have
    \(
    \zeta(\TpStraccia(\appint)) = \lpap(\zeta(\appint))
    \).
    \begin{longproof}
        We first show, by structural induction on the syntax of normal fuzzy formulas $\varphi$, that
        \(
        \fst{\fformeval{\appint}{\varphi}}=\fnnformeval{\zeta(\appint)}{\varphi}
        \).
        For the base cases, we have
        $\fst{\fformeval{\appint}{\atom}}
            =\fst{\appint(\atom)}
            =\lint(\atom)
            =\fnnformeval{\zeta(\appint)}{\atom}$, as well as
        $\fst{\fformeval{\appint}{\flpnot\atom}}
            =\fst{(\flpcontfStraccia{\flpnot}\fformeval{\appint}{\atom})}
            =\fst{(\flpcontfStraccia{\flpnot}\appint(\atom))}
            =\fst{(\flpcontfStraccia{\flpnot}\!\pair{\lint(\atom)}{\uint(\atom)})}
            =\fst{\pair{\flpnotf\uint(\atom)}{\flpnotf\lint(\atom)}}
            =\flpnotf\uint(\atom)
            =\fnformeval{\lint}{\uint}{\flpnot\atom}
            =\fnnformeval{\zeta(\appint)}{\flpnot\atom}$.
        The inductive cases are clear from the definition.
        Given $\appint$ with $\zeta(\appint)=(\lint,\uint)$, we denote by $\rappint$ the approximate interpretation with $\zeta(\rappint)=(\uint,\lint)$.
        For the main proof, we now find
        $\zeta(\TpStraccia(\appint))
            =\zeta\!\left(\set{\atom\mapsto\fformeval{\appint}{\body_\atom}\guard\atom\in\flpatom}\right)
            =\left(\set{\atom\mapsto\fst{\fformeval{\appint}{\body_\atom}}\guard\atom\in\flpatom},\set{\atom\mapsto\snd{\fformeval{\appint}{\body_\atom}}\guard\atom\in\flpatom}\right)
            =\left(\set{\atom\mapsto\fst{\fformeval{\appint}{\body_\atom}}\guard\atom\in\flpatom},\set{\atom\mapsto\fst{\fformeval{\rappint}{\body_\atom}}\guard\atom\in\flpatom}\right)
            =\left(\set{\atom\mapsto\fnnformeval{\zeta(\appint)}{\body_\atom}\guard\atom\in\flpatom},\set{\atom\mapsto\fnnformeval{\zeta(\rappint)}{\body_\atom}\guard\atom\in\flpatom}\right)
            =(\fst{\lpap(\lint,\uint)},\fst{\lpap(\uint,\lint)})
            =\lpap(\lint,\uint)
            =\lpap(\zeta(\appint))$.
        \hfill $\square$
    \end{longproof}
\end{lemma}

This result paves the way for our subsequent formal comparison of the well-founded semantics defined by \citet{LoyerS09} and the well-founded semantics given by approximation fixpoint theory~\cite{denecker00approximations}.
\newcommand{\closedworldProg}{\SpStraccia_\elp}
\newcommand{\falseint}{\appint_{\tt f}}
\newcommand{\bottomin}{\appint_\bot}
\newcommand{\stracciawellFoundedProg}{\WFStraccia_\elp}
The next definition makes use of two special interpretations:
$\falseint$ is such that $\zeta(\falseint)=\pair{\leastint}{\leastint}$ (it assigns false everywhere), and
$\bottomin$ is such that $\zeta(\bottomin)=\pair{\leastint}{\grtstint}$ (it assigns unknown everywhere).
The \define{closed world operator} $\closedworldProg$ is used to construct the well-founded semantics.
We will use the following characterization \cite[Theorem 4]{LoyerS09}:
Given an approximate interpretation $\appint$,  $\closedworldProg(\appint)$ is defined as the least fixpoint of the function
\(
\FpStraccia[\elp]{\appint}(\appintalt)\eqdef\falseint\pglb\TpStraccia(\appint\plub\appintalt)
\),
that is,
$\SpStraccia_\elp(\appint)\eqdef\lfp_{\tleq}(\FpStraccia[\elp]{\appint})$.
The \define{approximate well-founded operator} is then defined as:
\(
\stracciawellFoundedProg(\appint) = \TpStraccia(\appint)\plub\closedworldProg(\appint)
\).
Finally, the \define{approximate well-founded model} is
$\lfp_{\pleq}(\WFStraccia_\elp)$.

\begin{lemma}\label{lemma:closedworld:is:stable}
    For any \flptext $\flp$ and any approximate interpretation $\appint$ over $\flpatom$ with \mbox{$\zeta(\appint)=(\lint,\uint)$}, we have that \mbox{$\zeta(\closedworldProg(\appint))=(\leastint,\lfp(\snd{\lpap(\lint,\cdot)}))$}.
    \begin{longproof}
        In what follows, for an approximate interpretation $\appint$, we denote with $\lint$ respectively $\uint$ the interpretations such that $\zeta(\appint)=(\lint,\uint)$.

        We first note that, ($\dagger$):
        for any partial interpretation $\appint$ and any $p\in \flpatom$, $\zeta(\falseint\pglb\appint)(p)=(\least, \uint(p))$
        or equivalently, $\zeta(\falseint\pglb\appint)=(\leastint,\uint)$.
        Likewise, ($\ddagger$):
        for any partial interpretation $\appint$ and any $p\in \flpatom$, $\zeta(\appint\plub\falseint)(p)=(\lint(p),\least)$, or equivalently, $\zeta(\appint\plub\falseint)=(\lint, \leastint)$.

        We now show by induction on the construction of $\SpStraccia_\elp(\appint)=\lfp_{\tleq}(\FpStraccia[\elp]{\appint})$ that  $  \zeta(\closedworldProg(\appint))=(\leastint,\lfp(\snd{\lpap(\lint,\cdot)}))$.

        For the base case, we have to show that $\zeta(\FpStraccia[\elp]{\appint}(\falseint))=(\leastint, \snd{\lpap(\lint,\leastint)})$.
        Indeed, notice first that $ \FpStraccia[\elp]{\appint}(\falseint)=\falseint\pglb\TpStraccia(\appint\plub\falseint)$ by definition.
        With ($\ddagger$), $\zeta(\appint\plub\falseint)=(\lint,\leastint)$.
        With Lemma \ref{lem:straccia:operators-isomorphic}, $ \zeta(\TpStraccia(\appint\plub\falseint))=\lpap(\lint,\leastint)$.
        With ($\dagger$), $\falseint\pglb\TpStraccia(\appint\plub\falseint)= (\leastint, \snd{\lpap(\lint,\leastint)})$, thus showing the base case.

        The inductive case is similar. %
        \hfill $\Box$
    \end{longproof}
\end{lemma}

We now show that the approximate well-founded model is identical to the well-founded fixpoint of $\lpap$.
Intuitively, we prove this by relating the paths that the operators $\WFStraccia_\elp$ and $\stable{\lpap}$ traverse when applied iteratively to $\bottomin\mathbin{\hat=}\pair{\leastint}{\grtstint}$:
every pair visited by $\stable{\lpap}$ is also visited by $\WFStraccia_\elp$, but not necessarily vice versa.
(This can also be seen in \Cref{fig:bilattice2}.) %
\begin{theorem}
    \label{thm:wf:main}
    For any \flptext $\flp$ with approximate well-founded model $\appint_{\sf SWF}$, we have $\zeta(\appint_{\sf SWF})=\lfp(\stable{\lpap})$.
    \begin{proof}
        In what follows, we will denote the sequence used to construct $\lfp_{\pleq}( \WFStraccia_\elp)$ by   $\langle \appint_i \rangle_{i\leq \alpha}$, and $\zeta(\appint_i)$ by $(\lint_i, \uint_i)$ (for any $i\leq \alpha$).\footnote{Notice that \citet{LoyerS09} do not define the approximate well-founded semantics in a constructive way, but refer to \citet{tarski55fixpoint} to guarantee the existence of a least fixpoint. However, as explained in \Cref{subsec:aft} we can (without loss of generality) assume that this least fixpoint was constructed using an iterated (possibly transfinite) application of $\WFStraccia_\elp$.}
        We prove the theorem by showing that $\langle (\lint_i, \uint_i)\rangle_{i\leq \alpha}$ is a \define{terminal monotone induction} of $\stable{\lpap}$, i.e.\ a sequence $\langle (\lint_i, \uint_i)\rangle_{i\leq \alpha}$ such that:
        \begin{enumerate}
            \item $(\lint_0,\uint_0)=(\leastint,\grtstint)$,
            \item $(\lint_i, \uint_i)\plt(\lint_{i+1}, \uint_{i+1})\pleq\stable{\lpap}(\lint_i, \uint_i)$ for all $i<\alpha$,
            \item $(\lint_\sigma, \uint_\sigma)=\bigplub(\{(\lint_i, \uint_i)\mid i<\sigma\})$, for every limit ordinal $\sigma<\alpha$,
            \item there is no $(\lint_{\alpha+1},\lint_{\alpha+1})$ such that $\langle (\lint_i, \uint_i)\rangle_{i\leq \alpha+1}$ is a monotone induction.
        \end{enumerate}
        It then follows with \cite[Proposition 1]{denecker2007well} that $\zeta(\lfp_{\pleq}(\WFStraccia_\elp))=\lfp(\stable{\lpap})$.%
        We denote $\lfp(\snd{\lpap(\lint,\cdot)})$ by $\complete{\lpap}(\lint)$ to avoid clutter in what follows.

        Ad 1: Clear from the definition of $\zeta(\lfp_{\pleq}(\WFStraccia_\elp))$.

        Ad 2:
        We first show  by induction that ($\dagger$) $\langle (\lint_{i+1}, \uint_{i+1})\rangle= (\fst{\lpap(\lint_i, \uint_i)}, \complete{\lpap}(\lint_i))$.
        The base case is trivial. For the inductive case, suppose that  $\langle (\lint_{i+1}, \uint_{i+1})\rangle= (\fst{\lpap(\lint_i, \uint_i)},  \complete{\lpap}(\lint_i))$. We show that  $\langle (\lint_{i+2}, \uint_{i+2})\rangle= (\fst{\lpap(\lint_{i+1}, \uint_{i+1})},  \complete{\lpap}(\lint_{i+1}))$.
        Indeed as $\WFStraccia_\elp(\appint_{i+2})=\TpStraccia(\appint_{i+1})\pglb\closedworldProg(\appint_{i+1})$ and $\zeta(\closedworldProg(\appint_{i+1}))=(\least,\complete{\lpap}(\lint_{i+1}))$ (Lemma \ref{lemma:closedworld:is:stable}),
        it is easy to see that $\lint_{i+2}=\fst{\lpap(\lint_{i+1},\uint_{i+1})}$. It remains to show that $\uint_{i+2}=\complete{\lpap}(\lint_{i+1})$.
        We do this by showing that $\complete{\lpap}(\lint_{i+1})\lleq \lpap(\lint_{i+1},\uint_{i+1})$.
        As $\lint_i\lleq \lint_{i+1}$ (as  $\WFStraccia_\elp$ is a $\pleq$-monotone operator \cite[Theorem 7]{LoyerS09}) and $\stable{\lpap}$ is $\pleq$-monotone (which means that $\complete{\lpap}$ is $\lleq$-anti monotone),
        $\complete{\lpap}(\lint_{i+1})\lleq \complete{\lpap}(\lint_{i})= \uint_{i+1}$.
        As $\lpap$ is $\pleq$-monotone, $\snd{\lpap(\lint_{i+1},\cdot)}$ is $\lleq$-monotone, and thus,
        $\snd{\lpap(\lint_{i+1},\complete{\lpap}(\lint_{i+1}))}\leq \snd{\lpap(\lint_{i+1},\uint_{i+1})}$.
        As $\snd{\lpap(\lint_{i+1},\complete{\lpap}(\lint_{i+1}))}=\complete{\lpap}(\lint_{i+1})$ (after all, $\complete{\lpap}(\lint_{i+1})$ is the least fixed point of $\complete{\lpap}(\lint_{i+1},\cdot)$, this concludes the proof of $\dagger$.

        Back to the proof of the main claim. That $\appint_{i}\plt\appint_{i+1}$ follows immediately from that fact that $\WFStraccia_\elp$ is a $\pleq$-monotone operator \cite[Theorem 7]{LoyerS09}.
        We show that $\zeta(\TpStraccia(\appint_i))\pleq \stable{\lpap}(\zeta(\appint_i))$, by induction on $i$. The base case is clear. For the inductive case,
        suppose that $\appint_i \pleq \stable{\lpap}(\zeta(\appint_{i-1}))$.
        As $\appint_{i-1}\pleq \appint_i$ (as we just showed), and $\stable{\lpap}$ is a $\plt$-monotone operator, $\stable{\lpap}(\zeta(\appint_{i-1}))\pleq \stable{\lpap}(\zeta(\appint_i))$, and thus we have that  $\zeta(\appint_i )\pleq\stable{\lpap}(\zeta(\appint_i))$. This implies that $\lint_i\lleq \fst{\stable{\lpap}(\appint_i)}$.
        As $\fst{\lpap(\cdot, \uint_i )}$ is $\lleq$-monotone, $\fst{\lpap(\lint_i,\uint_i)}\lleq  \fst{\lpap( \fst{\stable{\lpap}(\appint_i)},\uint_i)}$. As $\fst{\stable{\lpap}(\appint_i)}$ is the least fixpoint of $\fst{\lpap(\cdot, \uint_i )}$, $\fst{\lpap( \fst{\stable{\lpap}(\appint_i)},\uint_i)}=\fst{\stable{\lpap}(\appint_i)}$. We infer that $\fst{\lpap(\lint_i,\uint_i)}\lleq \fst{\stable{\lpap}(\appint_i)}$, concluding the proof.

        Ad 3: This follows immediately from the definition of the iterated application of an operator.

        Ad 4: This is clear as we take the least fixpoint of $\zeta(\WFStraccia_\elp)$, which means that for any extension of the sequence, the condition $\appint_{i}\plt\appint_{i+1}$ is not satisfied.
        \hfill $\Box$
    \end{proof}
\end{theorem}

However, this result does not generalize to arbitrary partial stable interpretations (arbitrary fixpoints of $\stable{\lpap}$):%
\begin{example}\label{ex:partial:supported:aft:vs:straccia}
    Consider the normal fuzzy logic program \mbox{$\elp=\{ p\flpif{}_G q, p\flpif{}_G p, q\flpif{}_G \flpnot r, r\flpif{}_G \flpnot q\}$}.
    From \Cref{thm:wf:main}, it follows that the least fixpoints of $\stable{\lpap}$ and $\WFStraccia_\elp$ coincide (modulo isomorphism).
    However, the approximate interpretation
    \mbox{$\appint=\set{p\mapsto(1,1),q\mapsto(0,1),r\mapsto(0,1)}$} is a fixpoint of $\WFStraccia_\elp$, but $\zeta(\appint)$ is \emph{not} a fixpoint of $\stable{\lpap}$.
    Intuitively, $\WFStraccia_\elp$ cannot identify the positive cycle $p\flpif{}_G p$ that is needed to keep $p$ true in $\appint$, while $\stable{\lpap}$ \emph{can} identify it.
\end{example}

\begin{figure}[ht]
  \centering
  \begin{tikzpicture}[scale=0.6,node distance=8em, transform shape]
    \bilatticeEx
  \end{tikzpicture}
  \caption{%
    Parts of the bilattice of interpretations for \protect{\Cref{exm:main}}.
    The bilattice elements are represented by giving lower and upper truth value bounds for $p$, $q$, $r$, and $s$, respectively;
    so for example %
    \scalebox{0.7}{%
      \(%
      \arraycolsep=1.5pt
      \protect\renewcommand{\arraystretch}{0.8}
      \begin{array}{cccc}
        1.0 & 1.0 & 0.3 & 0.0 \\
        0.3 & 0.0 & 0.3 & 0.0
      \end{array}
      \)}
    represents the pair
    $\pair{\lint}{\uint}$ with lower bound
    $\lint=\set{p\mapsto0.3,q\mapsto0.0,r\mapsto0.3,s\mapsto0.0}$ and
    respective upper bound $\uint=\set{p\mapsto1.0,q\mapsto1.0,r\mapsto0.3,s\mapsto0.0}$.
    An operator $\op$ is visualized by an arrow from $\appint$ to $\op(\appint)$.
    Thus, the well-founded model of $\flp$ is found by
    starting from $(\leastint,\grtstint)$ and then following the $\stable{\lpap}$ (or $\WFStraccia_\elp$) arrows until the least fixpoint is reached.
  }
  \vspace*{-1ex}
  \label{fig:bilattice2}
\end{figure}

\section{Stable Model Semantics}
\label{sec:sm}

We now turn our attention to the stable model semantics for normal fuzzy logic programs as defined by \citet{cornejo2018syntax}.
As for classical (two-valued) normal logic programs, the notion of a stable model of a fuzzy normal logic program is associated with the least model of a positive program (whose consequence operator is monotone and thus has a unique least fixpoint).
Given a normal \flptext $\flp$ and an interpretation $\fint$, \citet{cornejo2018syntax} introduce a mechanism to obtain the positive program $\flp_{\fint}$ by substituting each rule \mbox{$r=\flpr$} in $\flp$ by a rule \mbox{$r_{\fint}\eqdef p\flpif{\flpweight}_i\body_{\fint}$}, where $\body_{\fint}$ is formally defined as follows:
Firstly, to indicate which atoms are used positively or negatively in the body formula $\body$, we write $\body[\atom_1,\ldots,\atom_m,\flpnot\atom_{m+1},\ldots,\flpnot\atom_{n}]$ as used by \citet{cornejo2018syntax}.
Secondly, every formula $\body[\atom_1,\ldots,\atom_n]$ of our language leads to an $n$-ary aggregator $\body[\cdot,\ldots,\cdot]$;
for example, the formula $\varphi[p,q,r]=p\land (q\lor r)$ leads to the aggregator $\varphi[\cdot,\cdot,\cdot]$ (that could also produce $\varphi[r,p,q]=r\land(p\lor q)$) with associated truth function $\flpcontf{\varphi}(x_1,x_2,x_3)=x_1\flpcontf{\land}(x_2\flpcontf{\lor}x_3)$.
This allows to define the associated truth function of $\body_{\fint}$ via\/:
\begin{center}
    \vspace*{-2.6ex}
    \mbox{$\flpcontf{\body_{\fint}}(x_1,\ldots,x_m) = \flpcontf{\body}\!(x_1,\ldots,x_m,\flpnotf\fint(\atom_{m+1}),\ldots,\flpnotf\fint(\atom_{n}))$}
    \vspace*{-2.6ex}
\end{center}
Intuitively, $\body_{\fint}$ evaluates all and only the atoms that appear \emph{negatively} in $\body$ by $\fint$, and keeps the atoms that occur positively.
The program $\flp_{\fint}\eqdef\set{r_{\fint}\guard r\in\flp}$ is the \define{reduct of $\flp$} w.r.t.~$\fint$.
An $\fint\in\Fint$ is then a \define{stable model} of a normal \flptext if and only if $\fint$ is the least model of $\flp_{\fint}$~\cite{cornejo2018syntax}.

In the remainder of this section, we show how to reconstruct the stable model definition of \citet{cornejo2018syntax} within approximation fixpoint theory.
The first result relates our approximator with \citeauthor{cornejo2018syntax}'s definition of reduct.
\begin{lemma}
    \label{lem:reduct-vs-stable-operator}
    For any \flptext $\flp$ and \mbox{$\fint\in\Fint$},\; \mbox{$\fst{\lpap(\cdot,\fint)}=\lpop[\flp_{\fint}]$}.
    \begin{longproof}\footnote{The proof is similar to \cite[Proposition~26]{cornejo2018syntax}.}
        Given a formula $\body$, we have, for all $\fintalt,\fint\in\Fint$,
        \begin{align*}
            \fnformeval{\fintalt}{\fint}{\body} & = \fnformeval{\fintalt}{\fint}{\body[\atom_1,\ldots,\atom_m,\flpnot\atom_{m+1},\ldots,\flpnot\atom_{n}]}                  \\
                                                & = \flpcontf{\body}(\fintalt(\atom_1),\ldots,\fintalt(\atom_m),\flpnotf\fint(\atom_{m+1}),\ldots,\flpnotf\fint(\atom_{n})) \\
                                                & = \flpcontf{\body_{\fint}}(\fintalt(\atom_1),\ldots,\fintalt(\atom_m))                                                    \\
                                                & = \fformeval{\fintalt}{\body_{\fint}[\atom_1,\ldots,\atom_m]}                                                             \\
                                                & = \fformeval{\fintalt}{\body_{\fint}}
        \end{align*}
        Now, for $\fint\in\Fint$, we have $\flp_\fint=\set{\flprRed[\fint]\guard\flpr\in\flp}$ by definition and therefore
        for all $\fintalt\in\Fint$,
        \begin{align*}
            \fst{\lpap(\fintalt,\fint)}(\atom) & =
            \biglub \set{\flpweight\flpcontf{\flpand_i}\fnformeval{\fintalt}{\fint}{\body} \guard \flpr \in \flp}                                                   \\
                                               & = \biglub \set{\flpweight\flpcontf{\flpand_i}\fformeval{\fintalt}{\body_{\fint}} \guard \flprRed \in \flp_{\fint}} \\
                                               & = \lpop[\flp_{\fint}](\fintalt)(\atom)
            \hspace*{42mm}\Box
        \end{align*}
    \end{longproof}
\end{lemma}
It is clear that equal operators have equal fixpoints:
\begin{corollary}
    For any $\fint\in\Fint$,\; $\lfp(\fst{\lpap(\cdot,\fint)})=\lfp(\lpop[\flp_{\fint}])$.
\end{corollary}

This result now allows to prove the main result of this section, namely the correspondence of the two variants of definining the stable model semantics for fuzzy logic programs.

\begin{theorem}
    \label{thm:sm-correspondence}
    An exact interpretation $(\fint,\fint)$ is a (AFT) stable model of $\flp$ iff
    $\fint$ is a (\citeauthor{cornejo2018syntax}) stable model of $\flp$.
    \begin{proof}
        $(\fint,\fint)$ is an AFT stable model of $\flp$ iff
        $\stable{\lpap}(\fint,\fint)=(\fint,\fint)$ iff
        $\lfp(\fst{\lpap(\cdot,\fint)})=\fint$ iff
        $\lfp(\lpop[\flp_{\fint}])=\fint$ iff
        $\fint$ is a \citeauthor{cornejo2018syntax} stable model of $\flp$.
        \hfill $\Box$
    \end{proof}
\end{theorem}

\section{Applying AFT to Fuzzy Logic Programming}
\label{sec:aft4flp}

In this section, we derive several useful results that follow as straightforward corollaries from the fact that we used AFT as a unifying framework for representing the semantics of \flpstext.
In more detail, we investigate the relation between well-founded and stable model semantics (\Cref{subs:wf&stab}), provide results on stratification (\Cref{subs:stratification}), and introduce the ultimate semantics for fuzzy logic programming (\Cref{subsec:ultimate}).

\subsection{Relationship Between %
    Semantics}\label{subs:wf&stab}
One of the benefits of the operator-based characterization given in this paper is that it allows us to straightforwardly derive results on the relationship between different semantics considered here.
Firstly we consider the fuzzy well-founded and stable model semantics.
In AFT, it is well-known that the well-founded fixpoint is an approximation of every exact stable fixpoint.
Combining this insight with the characterizations developed in this paper, we obtain the following result, relating the semantics of
\citet{LoyerS09} and \citet{cornejo2018syntax}.
\begin{theorem}
    For any (AFT) stable model $(I,I)$ of $\flp$, we have \mbox{$\lfp(\stable{\lpap})\pleq (I,I)$}.
    Equivalently, for any (\citeauthor{cornejo2018syntax}) stable model $I$ of $\flp$, we have \mbox{$\zeta(\appint_{\sf SWF})\pleq (I,I)$}.
\end{theorem}

Secondly, we show that the fuzzy semantics are generalizations of the classical, two-valued semantics.
This is not obvious from their original definitions in the literature, but becomes obvious after our reconstruction in AFT.
\begin{proposition}
    Assume the complete lattice of truth values \mbox{$\lset=\set{\least,\grtst}$} with \mbox{$\least\llt\grtst$}.
    For every classical normal logic program $\flp$:
    (1) its well-founded model according to \citet{LoyerS09} coincides with its well-founded model according to \citet{GelderRS91};
    (2) its stable models according to \citet{cornejo2018syntax} coincide one-to-one with its stable models according to \citet{GelfondL88}.
\end{proposition}

\subsection{Stratification for Fuzzy Logic Programs}\label{subs:stratification}
We first recall some necessary preliminaries \cite{vennekens2006splitting}, adapted to our setting.\footnote{In the general algebraic setting of \citet{vennekens2006splitting}, these definitions are introduced using the notion of a product lattice, but we could simplify these notions as done above.}
Given an interpretation $\fint\in \Fint$, and a set of atoms $\Atoms'\subseteq \Atoms$, $\fint_{|\Atoms'}: \Atoms'\rightarrow \lset$ is defined by $\fint_{|\Atoms'}(p)=\fint(p)$ for any $p\in\Atoms'$.
For elements $(\lint,\uint)$ of the bilattice $\bil$, we denote $(\lint_{|\Atoms'},\uint_{|\Atoms'})$ by $(\lint,\uint)_{|\Atoms'}$.
The order $\lleq$ can likewise be restricted, denoted by $\lleq_{|\Atoms'}$.
Given an operator $\op$ over $\Fint$ relative to a set of atoms $\Atoms$, where $\Atoms_1,\Atoms_2$ forms a partition of $\Atoms$,
$\op$ is \emph{stratifiable} over $\Atoms_1,\Atoms_2$ iff for every $\fint^1,\fint^2\in \Fint$, if $\fint^1_{|\Atoms_1}=\fint^2_{|\Atoms_1}$ then $\op(\fint^1)_{|\Atoms_1}=\op(\fint^2)_{|\Atoms_1}$.
Some of the main results of \citet{vennekens2006splitting} include that the the construction of (or search for) all major AFT~semantics can be split up along the strata an operator is stratifiable over.

We now recall a syntactical criterion for identifying strata based on dependency graphs, again inspired by the work of \citet{vennekens2006splitting}.
This criterion generalizes the well-known definition known from normal logic programs \cite{AptBW88,vanGelder88}.
We restrict attention to stratifications consisting of two strata for simplicity, but these results can be extended straightforwardly to arbitrary numbers of strata.
We first need some additional preliminaries.
Given a program $\flp$ and $p,q\in \Atoms$, we define $q\preceq_\flp p$ iff there is a $\flpr\in \flp$ such that $q\in \body$.
We then say that $\flp$ is \emph{stratifiable} over the partition $\Atoms_1,\Atoms_2$ of $\Atoms$ iff $q\preceq_\flp p$, $q\in\Atoms_i$, and $p\in\Atoms_j$ imply that $i\leq j$.
Informally, if two atoms depend on each other, they either occur in the same stratum or the dependent atom occurs in the higher stratum.
This is an immediate generalization (at least for the case of two strata) from normal to fuzzy logic programs of the definition of splitting by \citet{vennekens2006splitting}.
The traditional notion of stratification \cite{AptBW88,vanGelder88} additionally requires that %
negatively occuring body atoms come from a \emph{strictly lower} stratum.

\begin{theorem}
    For any \flptext $\flp$, if $\flp$ is stratifiable over $\Atoms_1,\Atoms_2$, then $\lpap$ is stratifiable over $\Atoms_1,\Atoms_2$.
    \begin{proof}[Sketch]
        Suppose that $\flp$ is stratifiable over $\Atoms_1,\Atoms_2$.
        We show that $(\dagger)$: for any $(\lint^1,\uint^1),(\lint^2,\uint^2)\in \bil$,
        if $(\lint^1_{|\Atoms_1},\uint^1_{|\Atoms_1})=(\lint^2_{|\Atoms_1},\uint^2_{|\Atoms_1})$, then $\lpap(\lint^1,\uint^1)_{|\Atoms_1}= \lpap(\lint^2,\uint^2)_{|\Atoms_1}$.
        Consider some $p\in \Atoms_1$, some $\flpr\in \flp$ and some $q\in \body$.
        Thus $q\preceq_\flp p$ and since $\flp$ is stratifiable, $p\in\Atoms_1$ implies $q\in\Atoms_1$.
        Thus, we have established that $p\in\Atoms_1$ implies $q\in \Atoms_1$, for any $\flpr\in \flp$ and  $q\in \body$.
        $(\dagger)$ now follows immediately from the definition of $\lpap$.
        \hfill $\Box$
    \end{proof}
\end{theorem}
It follows now from the results of \citet{vennekens2006splitting}  (Theorem~3.11 and Corollary~3.12), that the construction of the well-founded fixpoint and the search for stable fixpoints can be split up along the strata formed in a dependency graph.
In practice, this is done by first doing the necessary computations for the lower stratum $\Atoms_1$, and then transforming the program as to take into account the computed values.
More technically, given an interpretation $(\lint^1,\uint^1)$ for $\Atoms_1$ and a formula $\body$, let $\body_{(\lint^1,\uint^1)}$ be
the formula obtained by replacing every positive $p\in\Atoms_1$ by $\lint^1(p)$ and every negative $\flpnot p$ by $\flpnotf\uint^1(p)$, and let  $\flp_{(\lint^1,\uint^1)}= \{ p \flpif{\flpweight}_i \body_{{(\lint^1,\uint^1)}}\mid \flpr\in \flp\}$.
\begin{lemma}
    Suppose that $\flp$ is stratifiable over $\Atoms_1,\Atoms_2$. Given an interpretation $(\lint,\uint)$ over $\Atoms_1\cup\Atoms_2$ and an interpretation $(\lint^1,\uint^1)$ over $\Atoms_1$, if $(\lint,\uint)_{|\Atoms_1}=(\lint^1,\uint^1)$ then $\mathcal{T}_{\flp_{(\lint^1,\uint^1)}}(\lint,\uint)_{|\Atoms_2}=(\lpap(\lint,\uint))_{|\Atoms_2}$.
    \begin{longproof}
        Consider some $p\in \Atoms_2$.
        We show by induction on $\body$ that for every $\flpr\in\flp$,
        $\fnformeval{\lint}{\uint}{\body}=\fnformeval{\lint_{|\Atoms_2}}{\uint_{|\Atoms_2}}\body_{(\lint^1,\uint^1)}$, from which the lemma follows straightforwardly.

        For the base case we consider two cases:
        (a) $\body=\atom$. In that case, if $\atom\in \Sigma_1$ then $\body_{(\lint^1,\uint^1)}=\lint^1(\atom)=\lint(\atom)$. Otherwise, i.e.\ if $\atom\in \Sigma_2$, $\body_{(\lint^1,\uint^1)}=\atom$.
        (b) $\body=\flpnot \atom$. In that case, if $\atom\in\Sigma_1$ then $\body_{(\lint^1,\uint^1)}=\flpnotf\uint(\atom)=(\lint,\uint)(\body)$. Otherwise, i.e.\ if $\atom\in \Sigma_2$, $\body_{(\lint^1,\uint^1)}=\flpnot \atom$.

        For the inductive case, suppose the claim holds for $\varphi_1,\ldots,\varphi_n$ and suppose that $\body=f(\varphi^1,\ldots,\varphi^n)$. Then $\body_{(\lint^1,\uint^1)}=f( \varphi^1_{(\lint^1,\uint^1)},\ldots,\varphi^n_{(\lint^1,\uint^1)})$ and the claim follows from the inductive hypothesis.
        \hfill $\Box$
    \end{longproof}
\end{lemma}

\begin{theorem}
    For any \flptext $\flp$, if $\flp$ is stratifiable over $\Atoms_1,\Atoms_2$, then  $(\lint,\uint)$ is a stable [the well-founded] fixpoint of $\lpap$ iff $(\lint_{|\Atoms_1},\uint_{|\Atoms_1})$ is a stable [the well-founded] fixpoint of $(\lpap)_{|\Atoms_1}$ and $(\lint_{|\Atoms_2},\uint_{|\Atoms_2})$ is a stable [the well-founded] fixpoint of ${\mathcal T}_{\flp_{(\lint^1,\uint^1)}}$.
\end{theorem}

\subsection{Ultimate Semantics}\label{subsec:ultimate}

For a given operator \mbox{$\op\colon\lset\to\lset$}, there will typically be many different approximators $\ap$ for $\op$, and choosing among them might not be trivial.
\citet{denecker04ultimate} devised a way to construct the most precise approximator possible:
The \define{ultimate} approximator of $\op$ is given by \mbox{$\ultop\colon\bil\to\bil$} with \mbox{$\ultop(x,y)=\pair{\bigglb\op([x,y])}{\biglub\op([x,y])}$} where we denote \mbox{$\op([x,y])\eqdef\set{\op(z)\guard x\lleq z\lleq y}$}.\footnote{Most precise here means that for all approximators $\ap$ of $\op$ and pairs $\pair{x}{y}\in\bil$, it holds that $\ap(x,y)\pleq\ultop(x,y)$.}
The usual AFT definitions of semantics relying on fixpoints of $\stable{\ultop}$ then yield \define{ultimate} versions of the semantics, e.g.\ the \define{ultimate well-founded semantics} of $\ultop$ is given by $\lfp(\stable{\ultop})$.

The ultimate approximation of the operator $\lpop$ on normal \flpstext
(\Cref{sec:approximator})
is thus given by $\ult{{\lpop}}$, which we denote by $\ultlpop$ for ease of notation.
The following example illustrates the potential advantages of $\ultlpop$ in comparison to $\lpap$.

\begin{example}
    \label{exm:ultlpop}
    Consider for $\Atoms=\set{p}$ the fuzzy logic program
    \begin{narrowgather}
        \flp = \set{ \quad p\flpif{}_G p, \qquad p\flpif{}_G \lpnot p \quad }.
    \end{narrowgather}%
    While we have \mbox{$\lpap(\leastint,\grtstint)=(\leastint,\grtstint)$} for our approximator from \Cref{def:approximator},
    the ultimate approximator obtains the strictly better bounds \mbox{$\ultlpop(\leastint,\grtstint)=(\set{p\mapsto 0.5},\grtstint)=\lfp(\ultlpop)$}.
    Consequently, for the well-founded semantics, we obtain
    $\fst{\lpap(\leastint,\grtstint)}=\leastint$ and
    $\snd{\lpap(\leastint,\leastint)}=\grtstint$ whence $(\leastint,\grtstint)$ is the well-founded fixpoint of $\lpap$;
    in contrast, using the ultimate approximator we get
    $\fst{\ultlpop(\leastint,\grtstint)}=\set{p\mapsto 0.5}$ and
    $\snd{\ultlpop(\leastint,\leastint)}=\grtstint$ whence $(\set{p\mapsto 0.5},\grtstint)$ is the well-founded fixpoint of $\ultlpop$.
\end{example}

For the semantics characterized by least fixpoints, it is known in AFT that the ultimate approximator leads to the most precise version of the semantics~\cite[Theorem~5.2]{denecker04ultimate}.
For fuzzy logic programming, we thus get:
\begin{proposition}
    For any \flptext $\flp$,\; $\lfp(\stable{\lpap})\pleq\lfp(\stable{\ultlpop})$.
\end{proposition}

Similarly, for semantics characterized by exact fixpoints, ultimate approximators preserve all exact (stable) models of other approximators, and may in general add new models~\cite[Theorem~5.3]{denecker04ultimate}.

\begin{proposition}
    For any \flptext $\flp$ and $\fint\in\Fint$: %
    \begin{enumerate}
        \item If $(\fint,\fint)$ is a fixpoint of $\lpap$, it is a fixpoint of $\ultlpop$;
        \item if $(\fint,\fint)$ is a stable fixpoint of $\lpap$, it is a stable fixpoint of $\ultlpop$.
    \end{enumerate}
\end{proposition}

Thus, having the semantics reconstructed within AFT, we could not only easily define fuzzy logic programming semantics that are more precise than those in the literature \cite{LoyerS09,cornejo2018syntax}, but also got results on how they relate to the less precise semantics.

\section{Discussion and Conclusion}

In this paper, we have reconstructed two specific semantics for fuzzy logic programming within approximation fixpoint theory.
This reconstruction enabled several novel results on fuzzy logic programming:
It allowed us to
(1) establish a formal relationship between the two semantics,
(2) generalize stratification to the case of fuzzy logic programs, and
(3) define \emph{ultimate} (stable/well-founded) semantics, which provide more information than their ordinary counterparts.

In future work, we want to analyze what current restrictions we could lift while keeping the positive, unifying aspects of our reconstruction.
Firstly, we intend to allow for a more general syntax of fuzzy logic programs, thus moving towards
(A) \define{extended} logic programs, where both classical negation $\neg$ and default negation $\lpnot$ can be used in rules~\cite{GelfondL91}, where \citet{Cornejo2020extended} proposed a (fuzzy) stable model semantics;
and/or
(B) \define{nested} logic programs~\cite{LifschitzTT99}, where negation can occur freely in program bodies.
Next, we also want to study lifting our requirements on negation;
in particular, the condition that negations $\flpnotf$ must be involutions, so that we can also allow Gödel negation.
Finally, we are interested in the (ordinal) number of steps that are needed to reach the least fixpoint of relevant operators such as $\lpap$ and $\stable{\lpap}$.
We are fairly confident that for finite programs with only rational explicit truth values and connectives that stay within the rationals, the fixpoint can be reached within $\omega=\left|\N\right|$ steps.

\iflong\else\clearpage\fi

\newcommand{\projectname}[1]{#1}
\section*{Acknowledgements}

This work is partly supported by BMBF (Federal Ministry of Education and Research) in DAAD project 57616814 (\href{https://secai.org/}{SECAI, School of Embedded Composite AI}).
We also acknowledge funding from BMBF within projects \projectname{KIMEDS} (grant no.~GW0552B), \projectname{MEDGE} (grant no.~16ME0529), and \projectname{SEMECO} (grant no.~03ZU1210B).

Jesse Heyninck was partially supported by the project \projectname{LogicLM}: Combining Logic Programs with Language Model with fle number
NGF.1609.241.010 of the research programme NGF AiNed
XS Europa 2024-1 which is (partly) financed by the Dutch
Research Council (NWO).

We thank Bart Bogaerts for discussions on this paper, including suggesting the proof structure of Theorem \ref{thm:wf:main} (inspired by his own work \cite{DBLP:conf/birthday/0001C24}), and coming up with Example \ref{ex:partial:supported:aft:vs:straccia}.

\bibliographystyle{named}
\bibliography{references}

\end{document}